\documentclass[pra,showpacs]{revtex4}
\usepackage{amsmath,amssymb}
\usepackage[dvips]{graphicx}
\newcommand \beq{\begin{eqnarray}}
\newcommand \eeq{\end{eqnarray}}
\begin{document}

\title{Dynamical symmetry in  spinor Bose-Einstein condensates} 
\author{Shun Uchino,$^{1}$ Takaharu Otsuka,$^{1,2,3}$ and Masahito Ueda$^{4,5}$}
\affiliation{
$^{1}$Department of Physics, University of Tokyo, Tokyo 113-0033, Japan\\
$^{2}$Center for Nuclear Study, University of Tokyo, Tokyo 113-0033, Japan\\
$^{3}$RIKEN, Saitama 351-0198, Japan\\
$^{4}$Department of Physics, Tokyo Institute of Technology, Tokyo 152-8551, Japan\\        
$^{5}$ERATO Macroscopic Quantum Project, JST, Tokyo 113-8656, Japan}
\begin{abstract}
 We demonstrate that dynamical symmetry plays a crucial role in determining the structure of the eigenspectra
of spinor Bose-Einstein condensates (BECs). 
In particular, the eigenspectra of spin-1 and spin-2 BECs in the single-mode approximation are shown to be completely determined by dynamical symmetries, where
a spin-2 BEC corresponds to the $U(5)$ limit of the interacting boson model in nuclear physics. 
The eigenspectrum of a spin-3 BEC is solved analytically for a specific class of coupling constants, 
while it is shown that dynamical symmetry alone is not sufficient to determine the spectrum for arbitrary coupling constants.
We also study the low-lying eigenspectra of spin-1 and spin-2 BECs in the absence of external magnetic field,
and find, in particular, that the quasi-degenerate spectra emerge for antiferromagnetic and cyclic phases. 
This implies that these systems are highly susceptible to external perturbations and 
may undergo symmetry-breaking transitions to other states upon increasing the system's size.

\end{abstract}
\maketitle
\section{Introduction}
Eigenvalue problems that can be solved analytically give profound insight into quantum many-body systems and are also important to elucidate the nature of the problems.
In many of such cases, the solvability originates from the symmetry of a system. Here, the symmetry includes not only the space-time symmetry
but also dynamical symmetry which arises from the special properties of the forces. 
If a system has a dynamical symmetry, the eigenvalue problem can be solved algebraically and we can find the exact spectrum which reveals the structure of the system. 
For instance, collective spectra of a number of atomic nuclei can be understood from dynamical symmetries \cite{arima}.

The $U(n)$ symmetry in the $n$-dimensional harmonic-oscillator problem
and the $O(4)$ symmetry in the three-dimensional Coulomb problem are two well-known examples of dynamical symmetries \cite{iachello}.
Dynamical symmetries have played crucial roles in such diverse fields of physics as elementary particle physics \cite{gell-mann}, nuclear physics \cite{arima}, and molecular physics \cite{iachello2}.
In the context of Bose-Einstein condensates (BECs) of dilute atomic vapor, the $SU(1,1)$ symmetry in a harmonically-trapped two-dimensional
system was discovered by Pitaevskii and Rosch \cite{pitaevskii}. 
In the present paper, we analyze the dynamical symmetry of spinor BECs.

Mean-field theories of spinor BECs have previously been discussed for spin-1, 2, and 3 cases
in, e.g., Refs. \cite{ho,ohmi}, \cite{koashi,ciobanu,ueda}, and \cite{diener,santos,makela2}, respectively.
The ground-state phases of spin-1 BECs comprise ferromagnetic and polar phases. 
In addition to these two phases, the ground-state phases in a spin-2 BEC involves a cyclic phase
which can hold 1/3-quantum vortices \cite{semenoff,makela}.
A rich variety of phases have been predicted in spin-3 BECs \cite{diener,santos,makela2,barnett,yip}; however, their physical properties  
have yet to be fully investigated.

The exact eigenspectra and eigenstates have been obtained for spin-1 and spin-2 BECs \cite{koashi,ueda,law,pu,ho2}.
For the case of antiferromagnetic coupling, it is found that the ground state of a spin-1 BEC is a spin-singlet pair-boson condensate 
which is fragmented in the sense that more than one eigenvalue of the single-particle density operator is of the order of $N$ \cite{koashi,law,mueller}.
As a consequence, the number of atoms in each magnetic sublevel fluctuates violently between 1 and $N$ as the magnetization of a system increases \cite{ho2,mueller}.
For the case of a spin-2 BEC, on the other hand, the cyclic and antiferromagnetic phases are predicted to exhibit a spin-singlet trio-boson condensate, and a Meissner-like effect in the magnetic response, respectively\cite{koashi,ueda}.

In this paper, we develop a systematic method of representing the eigenspectra and eigenstates of a spinor BEC by exploiting the dynamical symmetry
of the system,
and show that the eigenspectra and eigenstates of spin-1 and spin-2 BECs are determined for the entire range of coupling constants
and those of spin-3 BEC are determined for a specific class of coupling constants. 
In particular, a spin-2 BEC corresponds to the $U(5)$ limit of the interacting boson model, which describes the collective properties of atomic nuclei \cite{arima}, 
and the $U(5)$ classification can be utilized for representing excitation spectra of a spin-2 BEC.
We examine the low-lying eigenspectra and eigenstates of spin-1 and spin-2 BECs in the absence of external magnetic field
and find, in particular, that the quasi-degenerate spectra emerge above the ground states of antiferromagnetic and cyclic phases.
This fact suggests that these ground states are highly susceptible to external perturbations and may undergo symmetry-breaking transitions to
other states as the size of the system increases.

This paper is organized as follows. Section I\hspace{-.1em}I presents the Hamiltonian of a spin-$f$ BEC in terms of spherical tensor operators.
Section I\hspace{-.1em}I\hspace{-.1em}I  briefly reviews elements of group theory which are relevant to later discussions. 
Sections I\hspace{-.1em}V , V, and V\hspace{-.1em}I use the formalism described in Sec. I\hspace{-.1em}I\hspace{-.1em}I to analyze spin-1, -2, and -3 BECs, respectively.
Section V\hspace{-.1em}I\hspace{-.1em}I investigates the low-lying eigenspectra and eigenstates of spin-1 and spin-2 BECs
in the absence of external magnetic field.
Section V\hspace{-.1em}I\hspace{-.1em}I\hspace{-.1em}I summarizes the main results of the present paper.

\section{Formulation of the problem}
We consider a system comprised of identical bosons with spin $f$ and mass $M$ interacting via s-wave contact interaction.
The spin-independent part of the interaction is by far the largest and determines the density distribution of the particles.
We shall therefore assume that all particles share a single spatial mode (single-mode approximation), and focus on the many-body 
spectrum of the spin state.

The spin-dependent part of the interaction Hamiltonian can, in general, be written as \\
\beq 
 V(\mathbf{x_1}-\mathbf{x_2})=\delta (\mathbf{x_1}-\mathbf{x_2})\sum_{F=0,2,...,2f}g_F \mathcal{P}_F, 
\eeq
where $\displaystyle g_F=4\pi\hbar^{2}a_F/M$ is the coupling constant with $\displaystyle a_F$ being the s-wave scattering length in the total spin $F$ channel,
and $\displaystyle\mathcal{P}_F$ is the projection operator that projects the wave function of a pair of interacting atoms into the total spin $F$ channel. 
The  corresponding second-quantized Hamiltonian is given by
\beq
\hat{V}&=&\sum_{F=0,2,...,2f}\frac{g_F}{2} \sum_{M=-F}^{F}\int  d\mathbf{x} 
<fmfn|FM><FM|fm^{'}fn^{'}>\hat{\Psi}_m^{\dagger}\hat{\Psi}_n^{\dagger}\hat{\Psi}_{m^{'}}\hat{\Psi}_{n^{'}},  \label{secondint}
\eeq
where $\displaystyle <fmfn|FM>$ is the Clebsch-Gordan coefficient, and $\displaystyle\hat{\Psi}_m(\mathbf{x})$ $(m=-f,-f+1,...,f)$ represents the annihilation operator of a boson at position 
$\displaystyle\mathbf{x}$ with magnetic quantum number $m$. 
In Eq. (\ref{secondint}), repeated indices $(m, m^{'}, n, n^{'})$ are assumed to be summed.
In the presence of a magnetic field, the Hamiltonian also includes the Zeeman term
\beq
-p\int d\mathbf{x}\hat{\Psi}_m^{\dagger} (\hat{F}^{z})_{mn}\hat{\Psi}_{n}, 
\eeq
where $\displaystyle \hat{F}^{z}$ is the $z$ component of the spin operator, 
and $p$ is the product of the gyromagnetic ratio and the external magnetic field which is assumed to be applied in the $z$ direction.
In the following discussions, we shall write the sum of $\hat{V}$ and the Zeeman term as $\hat{H}$ and simply call it the Hamiltonian.
In the single-mode approximation the field operator takes the form of 
\beq
\hat{\Psi}_m\simeq\phi(\mathbf{x})\hat{a}_m,
\eeq
where $\displaystyle \phi(\mathbf{x})$ is the spatial mode into which all bosons are assumed to condense, 
and $\displaystyle \hat{a}_m$ is the corresponding annihilation operator with $m$ being the magnetic quantum number. 

The dynamical symmetry of the Hamiltonian can be analyzed in a transparent manner in terms of spherical tensor operators.
The spherical tensor operators of rank $f$, $\displaystyle\hat{T}^{f}_{m}$, transform as irreducible tensors under rotation and satisfy 
the following commutation relations:
\beq
[\hat{F}_z, \hat{T}^{f}_{m}]=m\hat{T}^{f}_{m}, \label{com1}
\eeq
\beq
[\hat{F}_{\pm},\hat{T}^{f}_{m}]=\sqrt{f(f+1)-m(m\pm1)}\hat{T}^{f}_{m\pm 1}, \label{com2}
\eeq
where $\displaystyle \hat{F}_{z}=\sum_m m\hat{a}^{\dagger}_{m}\hat{a}_m$, and
$\displaystyle \hat{F}_{\pm}=\sum_m\sqrt{f(f+1)-m(m\pm 1)}\hat{a}^{\dagger}_{m\pm 1}\hat{a}_{m}$.

We introduce a hierarchy of spherical tensor operators and tensor products as follows. We first note that operators
\beq
\hat{a}^{\dagger}_{m}, \ \ \hat{\tilde{a}}_{m}\equiv (-1)^{f-m}\hat{a}_{-m}, \label{tilde}
\eeq
satisfy conditions (\ref{com1}) and (\ref{com2}) and are therefore spherical tensor operators of rank $f$. 
In terms of these operators, we can introduce spherical tensor products of rank $l$ as
\beq
[\hat{a}^{\dagger}\times\hat{a}^{\dagger}]^{l}_{\mu}\equiv \sum_{m,n} <fmfn|l\mu>\hat{a}^{\dagger}_m\hat{a}^{\dagger}_n. \label{8}
\eeq
We note that $\displaystyle [\hat{a}^{\dagger}\times\hat{a}^{\dagger}]^{l}_{\mu}$ ($\mu =l,l-1,\cdots,-l$)
satisfy conditions (\ref{com1}) and (\ref{com2}) with
$f=l$ and are therefore the spherical tensors of rank $l$.
Since the Hamiltonian is a scalar, it should be expressed in terms of scalar quantities which can be constructed from Eq. (\ref{8}), as
\beq
[[\hat{a}^{\dagger}\times\hat{a}^{\dagger}]^{l}\times[\hat{\tilde{a}}\times\hat{\tilde{a}}]^{l}]^{0}&\equiv &
\sum_{\mu ,\mu^{'}}[[\hat{a}^{\dagger}\times\hat{a}^{\dagger}]^{l}_{\mu}\times[\hat{\tilde{a}}\times\hat{\tilde{a}}]^{l}_{\mu^{'}}]^{0}_0 \nonumber\\
&=&\sum_{m,m^{'},n,n^{'},\mu,\mu^{'}}<l\mu l\mu^{'}|00><fmfm^{'}|l\mu><l\mu^{'}|fnfn^{'}>\hat{a}^{\dagger}_{m}\hat{a}^{\dagger}_{m^{'}}
\hat{\tilde{a}}_{n}\hat{\tilde{a}}_{n^{'}}. \label{nine}
\eeq
In fact, it can be shown that the Hamiltonian is expressed in terms of the spherical tensor products as
\beq
\hat{H}&=&\frac{1}{2\Omega}\sum_{F=0,2,...,2f} g_F \sum_{M=-F}^{F} <fmfn|FM><FM|fm^{'}fn^{'}>\hat{a}_m^{\dagger} \hat{a}_n^{\dagger} \hat{a}_{m^{'}} \hat{a}_{n^{'}}
- p(\hat{F}^{z})_{mn}\hat{a}^{\dagger}_m\hat{a}_n \nonumber\\
	   &=&\frac{1}{2\Omega}\sum_{F=0,2,...,2f} \sqrt{2F+1} g_F [[\hat{a}^{\dagger}\times \hat{a}^{\dagger}]^{F}\times [\hat{\tilde{a}}\times\hat{\tilde{a}}]^{F}]^{0}
-p\sqrt{\frac{f(f+1)(2f+1)}{3}}[\hat{a}^{\dagger}\times\hat{\tilde{a}}]^{1}_{0}, \label{int}
\eeq 
where $\Omega\equiv(\int d\mathbf{x}|\phi|^{4})^{-1}$ is an effective volume.
The last expression in Eq. (\ref{int}) can be obtained as follows.
We use $\displaystyle <FM|fmfn>=<F,-M|f,-m,f,-n>$ and $\displaystyle <FMFM^{'}|00>=\frac{(-1)^{F-M}}{\sqrt{2F+1}}\delta_{M,-M^{'}}$ together with Eq. (\ref{tilde})
to show that
\beq
<FM|fm^{'}fn^{'}>\hat{a}_{m^{'}}\hat{a}_{n^{'}}&=&(-1)^{m^{'}+n^{'}}<FM|f,-m^{'},f,-n^{'}>\hat{\tilde{a}}_{m^{'}}\hat{\tilde{a}}_{n^{'}}\nonumber\\
														 &=&(-1)^{m^{'}+n^{'}}<F,-M|fm^{'}fn^{'}>\hat{\tilde{a}}_{m^{'}}\hat{\tilde{a}}_{n^{'}}\nonumber\\
														 &=&\sqrt{2F+1}<F,M,F,-M|00><F,-M|fm^{'}fn^{'}>\hat{\tilde{a}}_{m^{'}}\hat{\tilde{a}}_{n^{'}}
\eeq
Substituting this into the first lines of Eq. (\ref{int}) and comparing the result with Eq. (\ref{nine}), we obtain the desired expression.
In rewriting the Zeeman term, we have used $\displaystyle <f,m,f,-m|10>=m(-1)^{f-m}\sqrt{\frac{3}{f(2f+1)(f+1)}}$.

The eigenvalue problem of Hamiltonian (\ref{int}) thus reduces to expressing it in terms of invariant quantities  
that represent the underlying dynamical symmetry of the system. The invariant quantities can be found by using group theory which is described in the following section. 
\section{Elements of group theory}
To make this paper self-contained, we briefly review some elements of group theory that are relevant to our analysis of spinor BECs \cite{iachello}.
The one-particle properties of a spin-$f$ BEC are described by the generators of the unitary group $U(2f+1)$, 
$\hat{a}^{\dagger}_m\hat{a}_n$, which obey the commutation relations 
\beq
[\hat{a}^{\dagger}_m\hat{a}_n,\hat{a}^{\dagger}_{\mu}\hat{a}_{\nu}]=\delta_{n\mu}\hat{a}^{\dagger}_m\hat{a}_{\nu}-\delta_{m\nu}\hat{a}^{\dagger}_{\mu}\hat{a}_n.
\eeq
In addition, Bose symmetry requires that the bases for a system of $N$-identical bosons constitute a totally symmetric irreducible representation.
\subsection{Racah form}
The Racah form provides an alternative representation for the generators of the unitary group $U(2f+1)$ and is suitable for describing the interaction 
Hamiltonian with rotational invariance.
The Racah form can be constructed by the replacement of the generator $\displaystyle \hat{a}^{\dagger}_m\hat{a}_n$ of the $U(2f+1)$ group
with its rotationally covariant form $\displaystyle [\hat{a}^{\dagger}\times\hat{\tilde{a}}]^{l}_{\mu}$ ($l=0, \cdots, 2f$).
The commutation relations of the latter are given by 
\begin{widetext}
\beq
\left[ [\hat{a}^{\dagger}\times\hat{\tilde{a}}]^{l}_m,[\hat{a}^{\dagger}\times\hat{\tilde{a}}]^{l^{'}}_{m^{'}}\right]
&=&\sum_{l^{''},m^{''}}\sqrt{(2l+1)(2l^{'}+1)}<lml^{'}m^{'}|l^{''}m^{''}> \nonumber \\
&&\times \begin{Bmatrix} l & l^{'} & l^{''} \\ f & f & f \end{Bmatrix}\left[ (-1)^{l^{''}}-(-1)^{l+l^{'}}\right]
[\hat{a}^{\dagger}\times\hat{\tilde{a}}]^{l^{''}}_{m^{''}}, 
\eeq
\end{widetext}
where $\displaystyle \begin{Bmatrix} l & l^{'} & l^{''} \\ f & f & f \end{Bmatrix}$ is the Wigner 6-$j$ symbol whose definition and 
fundamental properties are given in Appendix. 
From the commutation relations (12), we find that $\displaystyle [\hat{a}^{\dagger}\times\hat{\tilde{a}}]^{l}_{\mu}$ with $l=1,3, \cdots, 2f-1$
form a subgroup of the $U(2f+1)$ group which is referred to as the $SO(2f+1)$ group \cite{iachello}.

The number operator and the angular momentum operator are expressed in terms of spherical tensor products as
\beq
&\hat{N}&=\sqrt{2f+1}[\hat{a}^{\dagger}\times\hat{\tilde{a}}]^{0} \label{number}
\eeq
and
\beq
&\hat{F}_m&=\sqrt{\frac{f(f+1)(2f+1)}{3}}[\hat{a}^{\dagger}\times\hat{\tilde{a}}]^{1}_{m}, \label{F}
\eeq
respectively, where $\displaystyle \hat{F}_1=\hat{F}_{+}=\hat{F}_{x}+i\hat{F}_y, \hat{F}_{-1}=\hat{F}_{-}=\hat{F}_x -i\hat{F}_y$, and $\displaystyle \hat{F}_0 =\hat{F}_z$.

The eigenvalues of the spinor Hamiltonian are related to the invariants called Casimir operators that commute with all the generators of 
the group. The Casimir operators relevant to spinor BECs are those of the $U(2f+1)$, $SO(2f+1)$, and $SO(2)$ groups.

We first consider the linear Casimir operators of the $U(2f+1)$ and $SO(2)$ groups. 
The Casimir operator that commutes with all the generators of the $U(2f+1)$ group is the number operator: 
\beq
\hat{C}_1(U(2f+1))&\equiv &\sqrt{2f+1}[\hat{a}^{\dagger}\times\hat{\tilde{a}}]^{0} \nonumber \\
				  &=&\hat{N}. \label{c1un}
\eeq
The $SO(2)$ group represents the rotation about the $z$-axis, and hereby has only one generator, with its linear Casimir operator given by 
\beq
\hat{C}_1(SO(2))&\equiv &\sqrt{\frac{f(f+1)(2f+1)}{3}}[\hat{a}^{\dagger}\times\hat{\tilde{a}}]^{1}_{0} \nonumber \\ 
                &=&\hat{F}_z.\label{c1so2}
\eeq

We next consider the quadratic Casimir operators of the $U(2f+1)$ and $SO(2f+1)$ groups. 
The quadratic Casimir operators are constructed by taking scalar products of the two generators.
The scalar product of tensor operators $\hat{T}^l$ and $\hat{U}^{l}$ of rank $l$ is defined by 
\beq
\hat{T}^{l}\cdot\hat{U}^{l}\equiv (-1)^{l}\sqrt{2l+1}[\hat{T}^{l}\times\hat{U}^{l}]^{0}.
\eeq
The quadratic Casimir operator of the $U(2f+1)$ group is given by 
\beq
\hat{C}_2(U(2f+1))&\equiv&\sum_{l=0}^{2f}[\hat{a}^{\dagger}\times\hat{\tilde{a}}]^{l}\cdot[\hat{a}^{\dagger}\times\hat{\tilde{a}}]^{l} \nonumber \\
&=&\hat{N}(\hat{N}+2f). \label{u_3} \label{c2un} 
\eeq
The quadratic Casimir operator of the $SO(2f+1)$ group is given for $f=1$ by
\beq
\hat{C}_2(SO(3))\equiv \frac{f(f+1)(2f+1)}{3}[\hat{a}^{\dagger}\times\hat{\tilde{a}}]^{1}\cdot[\hat{a}^{\dagger}\times\hat{\tilde{a}}]^{1},
\eeq
and for $f\neq 1$ by
\beq 
\hat{C}_2(SO(2f+1))&\equiv& 2\sum_{l=1,3,...,2f-1}[\hat{a}^{\dagger}\times \hat{\tilde{a}}]^l\cdot[\hat{a}^{\dagger}\times\hat{\tilde{a}}]^l \nonumber \\
&=&\hat{N}(\hat{N}+2f-1)-(2f+1)[[\hat{a}^{\dagger}\times \hat{a}^{\dagger}]^0\times[\hat{\tilde{a}}\times\hat{\tilde{a}}]^0]^0, \label{c2son}
\eeq
where a numerical factor of 2 in Eq. (\ref{c2son}) is introduced for the sake of convenience.
\subsection{Branching problem and dynamical symmetry}
To uniquely characterize the basis set of the problem, we need to find a complete set of quantum numbers.
This is done by introducing a complete chain of subgroups:
\beq
G\supset G^{'}\supset G^{''}\supset \cdots .
\eeq
To find the complete chain of subgroups is the aim of the branching problem, which can be solved according to the well-established rules \cite{iachello}.
Since the interaction Hamiltonian of spinor BECs has the $SO(3)$ symmetry, we consider the following chain:
\beq
U(2f+1)\supset SO(2f+1)\supset \cdots\supset SO(3)\supset SO(2).\label{chain}
\eeq
We also note that, to characterize the states uniquely, 
we must find $2f+1$ quantum numbers in the totally symmetric representations of the $U(2f+1)$ group.

When the Hamiltonian is expressed in terms of the Casimir operators of a chain of groups only, 
the system is said to possess dynamical symmetry, and the problem can be solved algebraically \cite{iachello}.
In the following sections we will discuss the dynamical symmetry of spinor BECs.  

\section{Spin-1 BEC}
We first apply the method presented in Sec. I\hspace{-.1em}I\hspace{-.1em}I to the eigenvalue problem of a spin-1 BEC.
Throughout this paper we consider only the $s$-wave scattering.
The Bose symmetry then requires that the total spin of two colliding bosons be either 0 or 2.
The corresponding scalar quantities that appear in the interaction Hamiltonian can be obtained from Eqs. (\ref{c2un}), (\ref{c2son}),
and (\ref{rec1}) as
\beq
[[\hat{a}^{\dagger}\times \hat{a}^{\dagger}]^0\times[\hat{\tilde{a}}\times\hat{\tilde{a}}]^0]^0=\frac{1}{3}\hat{N}(\hat{N}+1)-\frac{1}{3}\hat{F}^2, \label{kumikae_1}
\eeq
\beq
[[\hat{a}^{\dagger}\times \hat{a}^{\dagger}]^2\times[\hat{\tilde{a}}\times\hat{\tilde{a}}]^2]^0=\frac{1}{3\sqrt{5}}(2\hat{N}^2-4\hat{N}+\hat{F}^2) \label{kumikae_2}.
\eeq
Substituting Eqs. (\ref{kumikae_1}) and (\ref{kumikae_2}) into Eq. (\ref{int}), we obtain 
\beq
\hat{H}&=&\frac{1}{2\Omega}\sum_{F=0,2}\sqrt{2F+1} g_F [[\hat{a}^{\dagger}\times \hat{a}^{\dagger}]^F\times[\hat{\tilde{a}}\times\hat{\tilde{a}}]^F]^0
          -\sqrt{2}p[\hat{a}^{\dagger}\times\hat{\tilde{a}}]^{1}_{0} \nonumber\\
       &=&\frac{1}{2\Omega}\left[c_0\hat{N}(\hat{N}-1)+c_1(\hat{F}^2-2\hat{N})\right] -p\hat{F}_z\nonumber \\
       &=&\frac{1}{2\Omega}\left[c_0\hat{N}(\hat{N}+2)+c_2\hat{N}+c_1\hat{F}^2\right] -p\hat{F}_z \label{spin1},
\eeq
where the coupling constants $\displaystyle c_i$ ($i=0, \cdots, 2$) are given by
\beq
c_0=\frac{1}{3}g_0+\frac{2}{3}g_2, \ c_1=-\frac{1}{3}g_0+\frac{1}{3}g_2, \ c_2=-\frac{1}{3}g_0-\frac{8}{3}g_2.\nonumber\\
\eeq
The last equality in Eq. (\ref{spin1}) shows that the Hamiltonian is expressed as a sum of the Casimir operators of
the $U(3)$, $SO(3)$, and $SO(2)$ groups. This means that the Hamiltonian comprises the Casimir operators of the chain of the groups
\beq
U(3)\supset SO(3)\supset SO(2).
\eeq 
Since the Hamiltonian is expressed in terms of the Casimir operators of the chain of the groups only, the eigenvalue problem can be solved algebraically.
In the spin-1 case, we must specify three quantum numbers to characterize the state uniquely, as mentioned in Sec. I\hspace{-.1em}I\hspace{-.1em}I.
In the present case, we can choose $N, F,$ and $F_z$ as the desired quantum numbers which arise from the $U(3),$ $SO(3),$ and $SO(2)$ groups, respectively; therefore,
we obtain the eigenstate as  
\beq
|N,F,F_z>.\label{state1}
\eeq  
The corresponding eigenvalue of the Hamiltonian is 
\beq
E&=&\frac{1}{2\Omega}[c_0 N(N-1)+c_1F(F+1)]-pF_z \nonumber\\
	     &=&\frac{1}{2\Omega}[c_0N(N+2)+c_2N+c_1F(F+1)]-pF_z ,\nonumber\\ \label{eigen1}
\eeq
where we use $SU(2)$ algebras in order to obtain the eigenvalue of $\hat{F}^2$ and the allowed values of $F$ are 
$F=N,N-2,N-4,...,1 \ \text{or} \ 0$ \cite{law}.
The eigenvalue problem of a spin-1 BEC is thus completely solved by the dynamical symmetry of the system.
It is worthwhile to note that while all the Casimir operators that appear in the Hamiltonian (\ref{spin1}) commute with each other, 
the Hamiltonian itself does not commute with the generators of $U(3)$ group. This holds true for higher-spin cases to be described later.
The present situation is thus quite different from the familiar examples of dynamical symmetries such as the $n$-dimensional harmonic-oscillator problem and
the three dimensional Coulomb problem; in either of the latter two cases, the corresponding Hamiltonian commutes with the generators of the underlying group ($U(n)$ or $O(4)$). 
\section{Spin-2 BEC}
We next consider the case of a spin-2 BEC. The Bose symmetry requires that the total spin of two colliding bosons be 0, 2, and 4. 
The corresponding scalar quantities in the interaction Hamiltonian take the following forms:
\beq
[[\hat{a}^{\dagger}\times \hat{a}^{\dagger}]^0\times[\hat{\tilde{a}}\times\hat{\tilde{a}}]^0]^0
=\frac{1}{5}\left( \hat{N}^2+3\hat{N}-\hat{C}_2(SO(5)) \right) ,
\eeq
\beq
[[\hat{a}^{\dagger}\times \hat{a}^{\dagger}]^2\times[\hat{\tilde{a}}\times\hat{\tilde{a}}]^2]^0
=\frac{1}{7\sqrt{5}}\left( 2\hat{N}^2-4\hat{N}-\hat{F}^2+2\hat{C}_2(SO(5)) \right) ,
\eeq
\beq
[[\hat{a}^{\dagger}\times \hat{a}^{\dagger}]^4\times[\hat{\tilde{a}}\times\hat{\tilde{a}}]^4]^0
=\frac{1}{7}\left( \frac{6}{5}\hat{N}^2-\frac{12}{5}\hat{N}+\frac{1}{3}\hat{F}^2-\frac{1}{5}\hat{C}_2(SO(5)) \right) .
\eeq
Substituting these relations into Eq. (\ref{int}), we rewrite the Hamiltonian as
\beq
\hat{H}&=&\frac{1}{2\Omega}\sum_{F=0,2,4}\sqrt{2F+1}g_F [[\hat{a}^{\dagger}\times \hat{a}^{\dagger}]^F\times[\hat{\tilde{a}}\times\hat{\tilde{a}}]^F]^0 
          -\sqrt{10}p[\hat{a}^{\dagger}\times\hat{\tilde{a}}]^{1}_{0}\nonumber\\
	   &=&\frac{1}{2\Omega}\left[d_0\hat{N}(\hat{N}-1)+d_1(\hat{F}^{2}-6\hat{N})+\frac{4d_2}{5}\hat{S}_+\hat{S}_-\right]-p\hat{F}_z \nonumber \\
	   &=&\frac{1}{2\Omega}\left[ d_3\hat{N}(\hat{N}+4)
	   +d_4\hat{N}+d_5\hat{C}_2(SO(5))
	   +d_1\hat{F}^{2}\right]-p\hat{F}_z \label{spin2}
\eeq
where $\displaystyle \hat{S}_{+}=\frac{\sqrt{5}}{2}[\hat{a}^{\dagger}\times \hat{a}^{\dagger}]^{0}, \hat{S}_{-}=\hat{S}_{+}^{\dagger}$, 
and the coupling constants $d_i$ ($i=0, \cdots, 5$) are given by
\beq
d_0&=&\frac{4}{7}g_2+\frac{3}{7}g_4, \ d_1=-\frac{1}{7}g_2+\frac{1}{7}g_4, \ d_2=g_0-\frac{10}{7}g_2+\frac{3}{7}g_4, \nonumber\\
d_3&=&\frac{1}{5}g_0+\frac{2}{7}g_2+\frac{18}{35}g_4, \ 
d_4=-\frac{1}{5}g_0-\frac{12}{7}g_2-\frac{108}{35}g_4,\nonumber\\
d_5&=&-\frac{1}{5}g_0+\frac{2}{7}g_2-\frac{3}{35}g_4.
\eeq
The last equality in Eq. (\ref{spin2}) shows that the Hamiltonian comprises the Casimir operator of the $U(5),$ $SO(5)$, $SO(3)$, and $SO(2)$ groups.
The same situation arises in the $U(5)$ limit of the interacting boson model in nuclear physics \cite{arima}.
As in the latter case, the Hamiltonian is expressed in terms of the Casimir operators of the chain of the groups
\beq
U(5)\supset SO(5)\supset SO(3)\supset SO(2),
\eeq
and we can therefore find exact solutions, solely because of dynamical symmetry.
In the spin-2 case, we need five quantum numbers to uniquely characterize the state as mentioned below Eq. (\ref{chain}).

As in the spin-1 case, the quantum numbers  
$N, F,$ and $F_z$ arise from the $U(5),$ $SO(3),$ and $SO(2)$ groups, respectively.  
To find the quantum number of the $SO(5)$ group, we consider the following commutation relations: 
\beq
[\hat{S}_z,\hat{S}_{\pm}]=\pm \hat{S}_{\pm}, \ \ [\hat{S}_{+},\hat{S}_{-}]=-2\hat{S}_z,
\eeq
where $\displaystyle \hat{S}_z=\frac{1}{2}\hat{N}+\frac{5}{4}$. These operators satisfy $SU(1,1)$ commutation relations \cite{ueda}, 
and the Casimir operator of these operators is expressed as  
\beq
\hat{S}^2\equiv -\hat{S}_+\hat{S}_{-}-\hat{S}_z+\hat{S}_z^{2}.
\eeq
The eigenvalues of $\hat{S}^2$ and $\hat{S}_z$ are given by 
\beq
\hat{S}^2|S,S_z >=S(S-1)|S,S_z >, \ \hat{S}_z|S,S_z >=S_z |S,S_z >,\nonumber\\
\eeq
where $\displaystyle S_z = \frac{1}{2}N+\frac{5}{4}$. It follows from the relation
\beq
\hat{S}_{\pm}|S,S_z >=\sqrt{S^2_z \pm S_z -S(S-1)}|S,S_z >
\eeq
that the eigenvalue of $S_z$ for a given $S$ takes on values
\beq
S_z =S,S+1,S+2,\cdots .
\eeq
We introduce a quantum number $\tau$, which denotes the number of particles that do not form spin-singlet pairs.
Thus, $S$ is expressed in terms of $\tau$ by 
\beq
S=\frac{1}{2}\tau +\frac{5}{4}
\eeq
and $N$ takes on values
\beq
N=\tau ,\tau +2 ,\tau +4 ,\cdots.
\eeq
Since we can substitute $N$ and $\tau$  for $S$ and $S_z$, the eigenstates can be described by $|S,S_z > \equiv|N,\tau >$.
We thus find that the Casimir operator of the $SO(5)$ group is expressed as  
\beq
\hat{C}_2(SO(5))&=&\hat{N}(\hat{N}+3)-4\hat{S}_+\hat{S}_{-} \nonumber\\
	             &=&4\hat{S}^2-\frac{5}{4}.
\eeq
The eigenvalue of $\hat{C}_2(SO(5))$ is then
\beq
\hat{C}_2(SO(5))|S,S_z >&=&\left[ 4S(S-1)-\frac{5}{4}\right] |S,S_z > \nonumber\\
                         &=&\tau (\tau +3)|N,\tau >. \label{tau_1}
\eeq
This means that the quantum number of the $SO(5)$ group is $\tau$ $(\tau=N,N-2,N-4,\cdots,1\ \text{or}\ 0)$.
We have thus determined four quantum numbers $N, \tau , F,$ and $ F_z$. However, yet another quantum number is needed to uniquely characterize the 
states. This additional quantum number is not directly related to the Casimir operators and is called the missing label \cite{iachello}. 
We can choose this quantum number as the number of spin-singlet trios of bosons \cite{ueda}.
Thus the eigenstate of spin-2 BECs is given by
\beq
|N,\tau ,n_{30},F,F_{z} >, \label{state2}
\eeq
where $n_{30}$ is the number of spin-singlet trios, $F=\lambda,\lambda+1,...,2\lambda-2,2\lambda$ with $\tau =3n_{30}+\lambda$. (Note that $2\lambda-1$ is missing.) 
The complete basis set (\ref{state2}) and the relationship between the quantum numbers were pointed out in Ref. \cite{soryushi,chacon}.
The corresponding exact eigenvalue of the Hamiltonian is given by
\beq
E&=&\frac{1}{2\Omega}\left[ d_0 N(N-1)+\frac{d_2}{5}(N-\tau )(N+\tau +3)+d_1\{ F(F+1)-6N\} \right] -pF_z \nonumber\\
         &=&\frac{1}{2\Omega}\left[ d_3N(N+4)+d_4N+d_5\tau (\tau +3)+d_1F(F+1)\right] -pF_z. \label{eigen2}
\eeq 
The eigenvalue problem of a spin-2 BEC has thus been solved completely by dynamical symmetry alone.
We note that the eigenspectrum is degenerate with respect to the missing label $n_{30}$.
\section{Spin-3 BEC}
We finally investigate the case of a spin-3 BEC. 
This is the case in which dynamical symmetry alone cannot, in general, solve the eigenvalue problem. 
We will explain the reason for this and determine a specific class of coupling constants for which the problem can be solved by 
dynamical symmetry alone. We begin by noting the following relations:
\beq
\begin{Bmatrix} 5 & 5 & 3 \\ 3 & 3 & 3 \end{Bmatrix}=\begin{Bmatrix} 1 & 1 & 3 \\ 3 & 3 & 3 \end{Bmatrix}
=\begin{Bmatrix} 1 & 5 & 3 \\ 3 & 3 & 3 \end{Bmatrix}=0.
\eeq 
This means that $\displaystyle [\hat{a}^{\dagger}\times\hat{\tilde{a}}]^{l=1}_{m}$ and $\displaystyle [\hat{a}^{\dagger}\times\hat{\tilde{a}}]^{l=5}_{m}$ 
form a subgroup of the $SO(7)$ group,
which is referred to as the exceptional group $G_2$ \cite{iachello}. 
The quadratic Casimir operator of the $G_2$ group is given by
\beq
\hat{C}_2(G_2)\equiv 2\sum_{l=1,5}[\hat{a}^{\dagger}\times \hat{\tilde{a}}]^l\cdot[\hat{a}^{\dagger}\times\hat{\tilde{a}}]^l .\label{g2}
\eeq
Thus we can consider the following chain of groups:
\beq
U(7)\supset SO(7)\supset G_2\supset SO(3)\supset SO(2).
\eeq
The interaction Hamiltonian of spin-3 BECs has four independent coupling constants corresponding to the number of channels of binary collisions
with total spin 0, 2, 4, and 6. 
Since the number of the quadratic Casimir operators in the chain of the groups is also four, 
one might expect that the Hamiltonian can be expressed in terms of the Casimir operators alone. 
Unfortunately, this is not the case. To show this, we note that the following identity can be derived from Eq. (\ref{rec3}):
\beq
[\hat{a}^{\dagger}\times \hat{\tilde{a}}]^{1}\cdot[\hat{a}^{\dagger}\times\hat{\tilde{a}}]^{1}
-2[\hat{a}^{\dagger}\times \hat{\tilde{a}}]^{3}\cdot[\hat{a}^{\dagger}\times\hat{\tilde{a}}]^{3}
+[\hat{a}^{\dagger}\times \hat{\tilde{a}}]^{5}\cdot[\hat{a}^{\dagger}\times\hat{\tilde{a}}]^{5}=0. \label{identity}
\eeq
From Eqs. (\ref{c2son}), (\ref{g2}), and (\ref{identity}), we obtain
\beq
\hat{C}_2(G_2)=\frac{2}{3}\hat{C}_2(SO(7)).
\eeq
Thus the number of independent quadratic Casimir operators is three rather than four, and dynamical symmetry is 
not sufficient to completely determine the exact eigenvalues. If, on the other hand, we restrict ourselves to a specific class of 
the coupling constants, dynamical symmetry determines the eigenspectrum. 
In fact, using Eqs. (\ref{c2un}), (\ref{c2son}), and (\ref{rec3}), we obtain the following relations:
\beq
[[\hat{a}^{\dagger}\times \hat{a}^{\dagger}]^0\times[\hat{\tilde{a}}\times\hat{\tilde{a}}]^0]^0
=\frac{1}{7}\left( \hat{N}^2+5\hat{N}-\hat{C}_2(SO(7)) \right) ,
\eeq
\beq
[[\hat{a}^{\dagger}\times \hat{a}^{\dagger}]^2\times[\hat{\tilde{a}}\times\hat{\tilde{a}}]^2]^0
=\frac{1}{\sqrt{5}}\left( -\frac{10}{7}\hat{N}+[\hat{a}^{\dagger}\times \hat{\tilde{a}}]^{2}\cdot[\hat{a}^{\dagger}\times\hat{\tilde{a}}]^{2}
+\frac{5}{21}\hat{C}_2(SO(7))-\frac{5}{84}\hat{F}^2 \right) ,
\eeq
\beq
[[\hat{a}^{\dagger}\times \hat{a}^{\dagger}]^4\times[\hat{\tilde{a}}\times\hat{\tilde{a}}]^4]^0
=\frac{1}{3}\left( \frac{6}{11}\hat{N}^2+\frac{96}{77}\hat{N}-\frac{18}{11}[\hat{a}^{\dagger}\times \hat{\tilde{a}}]^{2}\cdot[\hat{a}^{\dagger}\times\hat{\tilde{a}}]^{2}
-\frac{9}{77}\hat{C}_2(SO(7))+\frac{1}{154}\hat{F}^2 \right) ,
\eeq
\beq
[[\hat{a}^{\dagger}\times \hat{a}^{\dagger}]^6\times[\hat{\tilde{a}}\times\hat{\tilde{a}}]^6]^0
=\frac{1}{\sqrt{13}}\left( \frac{24}{77}\hat{N}^2-\frac{118}{77}\hat{N}+\frac{7}{11}[\hat{a}^{\dagger}\times \hat{\tilde{a}}]^{2}\cdot[\hat{a}^{\dagger}\times\hat{\tilde{a}}]^{2}
+\frac{5}{231}\hat{C}_2(SO(7))+\frac{7}{132}\hat{F}^2 \right).
\eeq
Using these relations, we can rewrite the Hamiltonian as
\beq
\hat{H}&=&\frac{1}{2\Omega}\sum_{F=0,2,4,6}\sqrt{2F+1}g_F [[\hat{a}^{\dagger}\times \hat{a}^{\dagger}]^F\times[\hat{\tilde{a}}\times\hat{\tilde{a}}]^F]^0 
          -2\sqrt{7}p[\hat{a}^{\dagger}\times\hat{\tilde{a}}]^{1}_{0}\nonumber\\
	   &=&\frac{1}{2\Omega}\left[e_0\hat{N}(\hat{N}-1)+e_1(\hat{F}^{2}-12\hat{N})+\frac{4e_2}{7}\hat{S}_+\hat{S}_- +e_3
[\hat{a}^{\dagger}\times\hat{a}^{\dagger}]^{2}\cdot[\hat{\tilde{a}}\times\hat{\tilde{a}}]^{2} 
\right]-p\hat{F}_z \nonumber \\
	   &=&\frac{1}{2\Omega}\left[ e_4\hat{N}(\hat{N}+6)
	   +e_5\hat{N}+e_6\hat{C}_2(SO(7))
	   +e_7\hat{F}^{2}+e_3 
[\hat{a}^{\dagger}\times \hat{\tilde{a}}]^{2}\cdot[\hat{a}^{\dagger}\times\hat{\tilde{a}}]^{2}
\right]-p\hat{F}_z, \label{hspin3}
\eeq
where $\displaystyle \hat{S}_{+}=\frac{\sqrt{7}}{2}[\hat{a}^{\dagger}\times \hat{a}^{\dagger}]^{0}, \hat{S}_{-}=\hat{S}_{+}^{\dagger}$, 
and the coupling constants $e_i$ ($i=0, \cdots, 7$) are given by
\beq
e_0&=&\frac{9}{11}g_4+\frac{2}{11}g_6, \ e_1=-\frac{1}{11}g_4+\frac{1}{11}g_6, \nonumber\\
e_2&=&g_0-\frac{21}{11}g_4+\frac{10}{11}g_6, \ 
e_3=g_2-\frac{18}{11}g_4+\frac{7}{11}g_6, \nonumber\\
e_4&=&\frac{1}{7}g_0+\frac{6}{11}g_4+\frac{24}{77}g_6, \
e_5=-\frac{1}{7}g_0-\frac{10}{7}g_2-\frac{156}{77}g_4-\frac{262}{77}g_6, \nonumber\\
e_6&=&-\frac{1}{7}g_0+\frac{5}{21}g_2-\frac{9}{77}g_4+\frac{5}{231}g_6, \nonumber\\
e_7&=&-\frac{5}{84}g_2+\frac{1}{154}g_4+\frac{7}{132}g_6. \label{spin3}
\eeq
From Eq. (\ref{hspin3}) we see that, if $\displaystyle e_3=0$, dynamical symmetry alone can solve the problem, because 
the Hamiltonian would then be expressed in terms of the Casimir operators of
the $U(7),$ $SO(7),$ $SO(3),$ and $SO(2)$ groups only. We shall henceforth discuss the problem in this situation.

As in the case of spin-1 and spin-2 BECs, the quantum numbers of $N, F,$ and $F_z$ arise from the $U(7),$ $SO(3),$ and $SO(2)$ groups.  
The quantum number of the $SO(7)$ group is determined from the $SU(1,1)$ algebra:
\beq
[\hat{S}_z,\hat{S}_{\pm}]=\pm \hat{S}_{\pm}, \ \ [\hat{S}_{+},\hat{S}_{-}]=-2\hat{S}_z,
\eeq
where $\displaystyle \hat{S}_z=\frac{1}{2}\hat{N}+\frac{7}{4}$. As in the spin-2 case, the Casimir operator of the $SO(7)$ group is written as
 \beq
\hat{C}_2(SO(7))&=&\hat{N}(\hat{N}+5)-4\hat{S}_+\hat{S}_{-} \nonumber\\
	             &=&4\hat{S}^2-\frac{21}{4}.
\eeq
The eigenvalue of $\hat{C}_2(SO(7))$ is 
\beq
\hat{C}_2(SO(7))|S,S_z >&=&\left[ 4S(S-1)-\frac{21}{4}\right] |S,S_z > \nonumber\\
                         &=&\tau (\tau +5)|N,\tau >. 
\eeq
This means that the quantum number of the $SO(7)$ group is $\tau$. As a result, we have determined four quantum numbers; however, seven quantum numbers are needed to uniquely
characterize the state. Therefore, there must be three missing labels, and the eigenstate is expressed as
\beq
|N,\tau ,r,q,s,F,F_{z} > ,
\eeq 
where $r, q,$ and $s$ are missing labels \cite{roho}. The corresponding eigenvalue of the Hamiltonian is then given by
\beq
E_{e_3=0}&=&\frac{1}{2\Omega}\left[ e_0 N(N-1)+\frac{e_2}{7}(N-\tau )(N+\tau +5)+e_1\{ F(F+1)-12N\} \right] -pF_z \nonumber\\
         &=&\frac{1}{2\Omega}\left[ e_4N(N+6)+e_5N+e_6\tau (\tau +5)+e_7F(F+1)\right] -pF_z.
\eeq 
The eigenspectrum is degenerate with respect to missing labels $r,q,$ and $s$.

\section{Low-lying eigenspectra and eigenstates \label{low}}
In this section, we study the low-lying eigenspectra and eigenstates of spin-1 and spin-2 BECs
which are valid for an arbitrary range of coupling constants. 
\subsection{spin-1 case}
\begin{figure}
 \begin{center}
  \includegraphics{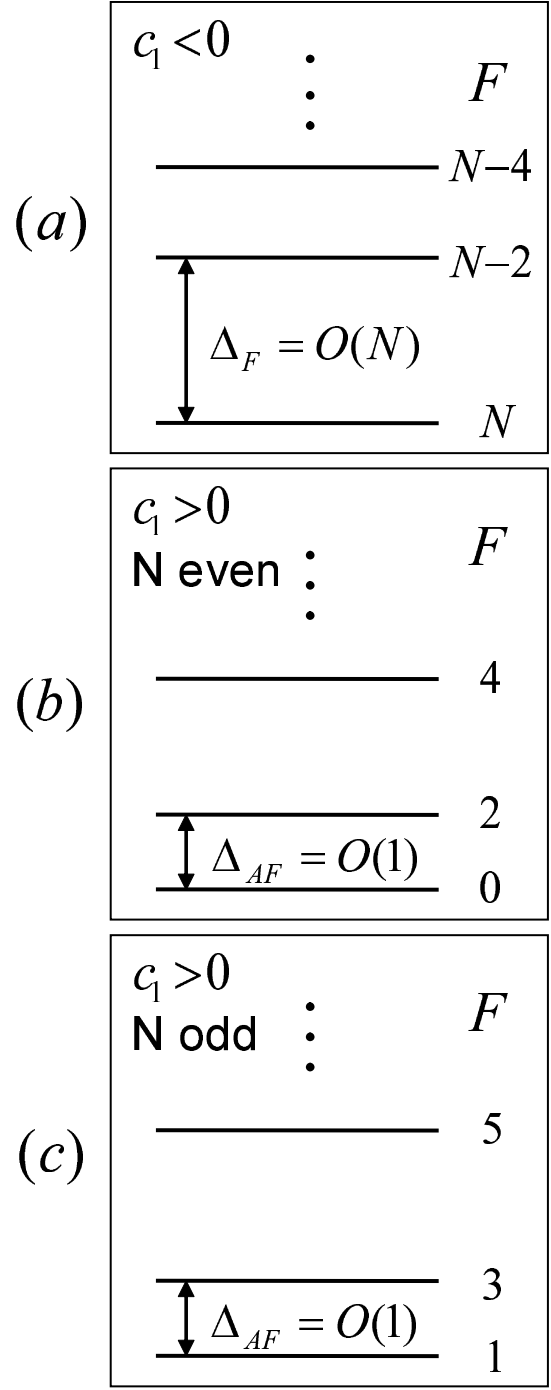}
  \caption{Low-lying eigenspectra of a spin-1 BEC. 
           The angular momentum $F$ of each state is shown to the right. (a) $c_1<0$. (b) $c_1>0$ with even $N$. (c) $c_1>0$ with odd $N$.
            The low-lying energy-level spacing is of the order of $N$ for the ferromagnetic BEC ($c_1<0$), while it is of the order of 1 for the antiferromagnetic BEC ($c_1>0$).
            $\Delta_{F(AF)}$ denotes the energy gap of the ferromagnetic (antiferromagnetic) ground state.}
  \label{Fig.1}
 \end{center}
\end{figure}

As can be seen from Eq. (\ref{eigen1}), the ground-state phase is determined by the sign of $c_1$ only;
it is ferromagnetic if $c_1<0$ and antiferromagnetic if $c_1>0$.
This result is consistent with the prediction of the mean-field theory.
The difference arises when the ground state is antiferromagnetic.
In this case, while the mean-field study predicts $F=0$, the exact result shows that $F$ is 0 or 1 
according to whether the number of particles $N$ is even or odd.
The low-lying excitation spectra show this even-odd parity effect as illustrated in Fig. \ref{Fig.1}.
The $F=1$ BEC of $^{87}$Rb is ferromagnetic ($c_1<0$) and that of $^{23}$Na is antiferromagnetic. 
Their low-lying spectra are respectively illustrated in Fig. 1 (a) and (b) or (c).
Since the ratio $\displaystyle \frac{c_1}{\tilde{g}}$ is of order $1$, 
where $\displaystyle\tilde{g}\equiv\frac{4\pi\hbar^2a_B}{M}$ and $a_B$ is the Bohr radius, 
a marked difference should arise in the low-lying energy-level spacing between the ferromagnetic and antiferromagnetic phases:
the low-lying energy-level spacing of the ferromagnetic BEC is of the order of $N$, but that of the antiferromagnetic BEC is of the order 1.
This implies that there are numerous quasi-degenerate energy levels above the ground state of the antiferromagnetic BEC, and the ground state is
therefore vulnerable to symmetry-breaking perturbations. 

\subsection{spin-2 case}
The exact ground-state phases and low-lying eigenspectra of a spin-2 BEC are much richer than those of a spin-1 BEC
since the number of the coupling constants
and quantum numbers increases from 1 to 2 and from 3 to 5, respectively.
The phase boundaries of the mean-field ground states of a spin-2 BEC have been shown as follows \cite{ciobanu,ueda}:
\beq
\text{ferromagnetic}: \ \  d_1<0 \ \ \text{and} \ \ d_5+4d_1<0, \\
\text{antiferromagnetic}: \ \  d_5>0 \ \ \text{and} \ \ d_5+4d_1>0, \\
\text{cyclic}: \ \ d_1>0 \ \ \text{and} \ \ d_5<0. 
\eeq
In many-body theory, from the analysis of (\ref{state2}) and (\ref{eigen2}), the ground-state phase boundaries are given as follows: 
\beq
\text{ferromagnetic}:&& d_1<0 \ \ \text{and} \ \ d_5+\frac{4N+2}{N+3}d_1<0 \ \ (N=2k,\ F=2N,\ \tau =N,\ n_{30}=0),\label{ferro1}\\
                     && d_1<0 \ \ \text{and} \ \ d_5+\frac{(2N-2)(2N+3)}{(N-1)(N+4)}d_1<0 \ \ (N=2k+1,\ F=2N,\ \tau =N,\ n_{30}=0),\label{ferro2}\nonumber\\ 
\eeq
\beq
\text{antiferromagnetic}:&&  d_5>0 \ \ \text{and} \ \ d_5+\frac{4N+2}{N+3}d_1>0 \ \ (N=2k,\ F=0,\ \tau =0,\ n_{30}=0), \\
             &&  d_5-\frac{3}{7}d_1>0 \ \ \text{and} \ \ d_5+\frac{(2N-2)(2N+3)}{(N-1)(N+4)}d_1>0 \ \ (N=2k+1,\ F=2,\ \tau =1,\ n_{30}=0),\label{antiferromagnetic2}\nonumber\\ \\
             &&  d_1>0 \ \ \text{and} \ \ 0<d_5<\frac{3}{7}d_1 \ \ (N=2k+1,\ F=0,\ \tau =3,\ n_{30}=1),\label{antiferromagnetic3}
\eeq
\beq
\text{cyclic}:&&  d_1>0 \ \ \text{and} \ \ d_5<0 \ \ (N=3k,\ F=0,\ \tau =N,\ n_{30}=\frac{N}{3}),\label{cyclic1}\\
              &&  d_1>0 \ \ \text{and} \ \ d_5+\frac{3}{4N-2}d_1<0 \ \ (N=3k+1,\ F=2,\ \tau =N,\ n_{30}=\frac{N-1}{3}), \label{cyclic2}\\
              &&  d_1>0 \ \ \text{and} \ \ -\frac{3}{4N-2}d_1<d_5<0 \ \ (N=3k+1,\ F=0,\ \tau =N-4,\ n_{30}=\frac{N-4}{3}),\label{cyclic3}\\
              &&  d_1>0 \ \ \text{and} \ \ d_5+\frac{3}{2N+1}d_1<0 \ \ (N=3k+2,\ F=2,\ \tau =N,\ n_{30}=\frac{N-2}{3}), \label{cyclic4}\\
              &&  d_1>0 \ \ \text{and} \ \ -\frac{3}{2N+1}d_1<d_5<0 \ \ (N=3k+2,\ F=0,\ \tau =N-2,\ n_{30}=\frac{N-2}{3}),\label{cyclic5}            
\eeq
where $k\in \mathbf{Z}$.
Thus, the phase boundaries and their number change compared with the case of mean-field theory
due to a finiteness of the number of particles.  
In fact, in the limit of $N\to \infty$, the phase boundaries between the ferromagnetic and antiferromagnetic phases reduce to those of mean-field theory.
If $N=6k$, the exact ground-state phases have a one-to-one correspondence to those of mean-field theory.
If we consider the cyclic phases with $N=3k+1$ or $N=3k+2$, the phases (\ref{cyclic3}) or (\ref{cyclic5}) disappear 
and the cyclic phases correspond to the phases (\ref{cyclic2}) or (\ref{cyclic4}) in this limit.
However, within the antiferromagnetic phases, the phase boundaries which appear with $N=2k+1$ do not disappear in this limit; thus, the antiferromagnetic phases
are always divided by (\ref{antiferromagnetic2}) and (\ref{antiferromagnetic3}).  

\begin{figure}
 \begin{center}
  \includegraphics{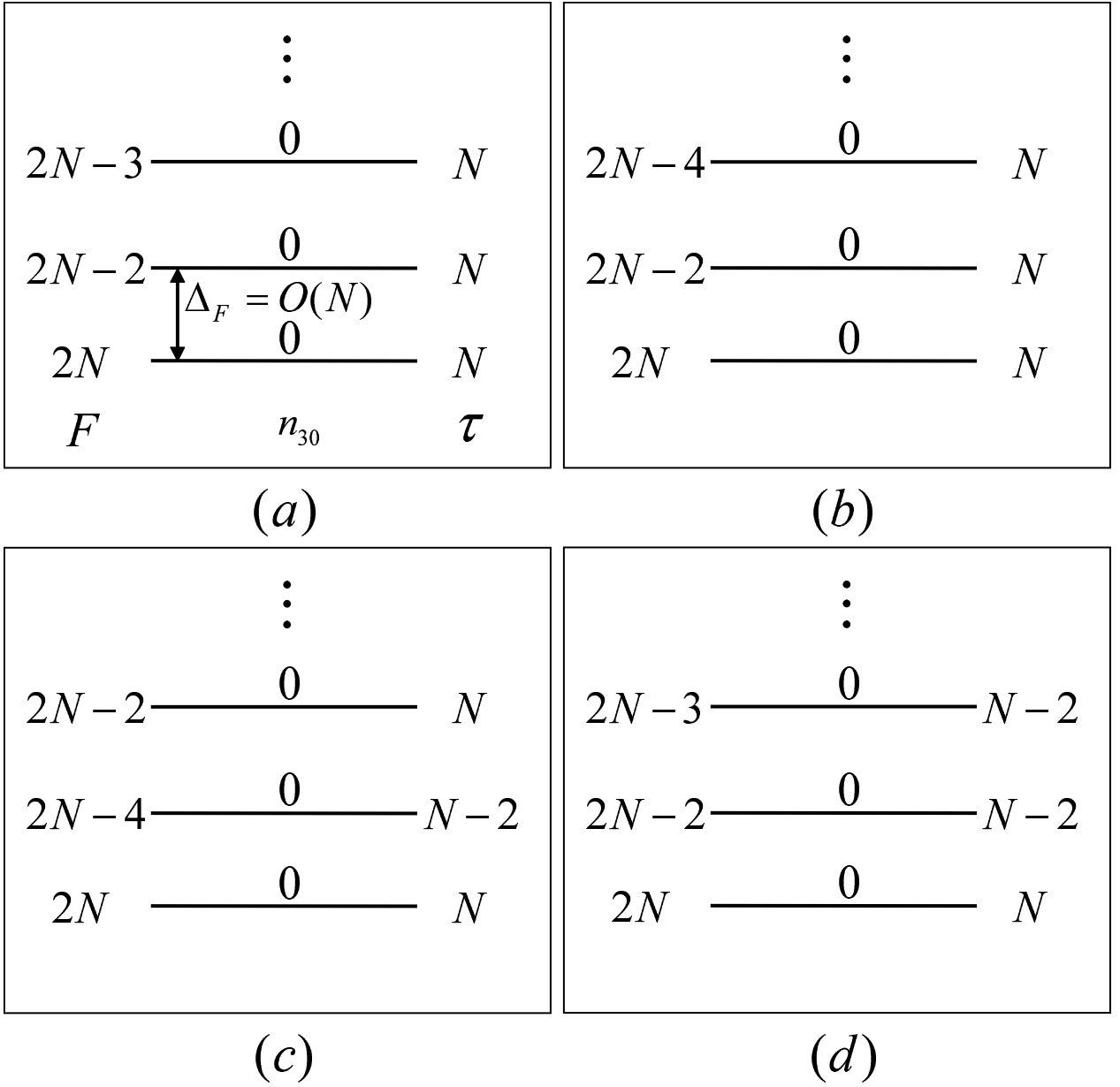}
  \caption{Low-lying eigenspectra of a spin-2 BEC for the ferromagnetic phases with $N=2k$ and $d_1<0$. 
           The values of $F$, $n_{30}$, and $\tau$ of each state are shown to the left, middle, and right, respectively.
		   (a) $d_5+\frac{2N-3}{2N+1}d_1<0$. (b) $-\frac{2N-3}{2N+1}d_1<d_5<-\frac{4N-5}{2N+1}d_1$.
		   (c) $-\frac{4N-5}{2N+1}d_1<d_5<-\frac{8N-14}{2N+1}d_1$. (d) $-\frac{8N-14}{2N+1}d_1<d_5<-\frac{4N+2}{N+3}d_1$.
			$\Delta_{F}$ denotes the energy gap of the ferromagnetic ground state.}
  \label{Fig.2}
 \end{center}
\end{figure}

\begin{figure}
 \begin{center}
  \includegraphics{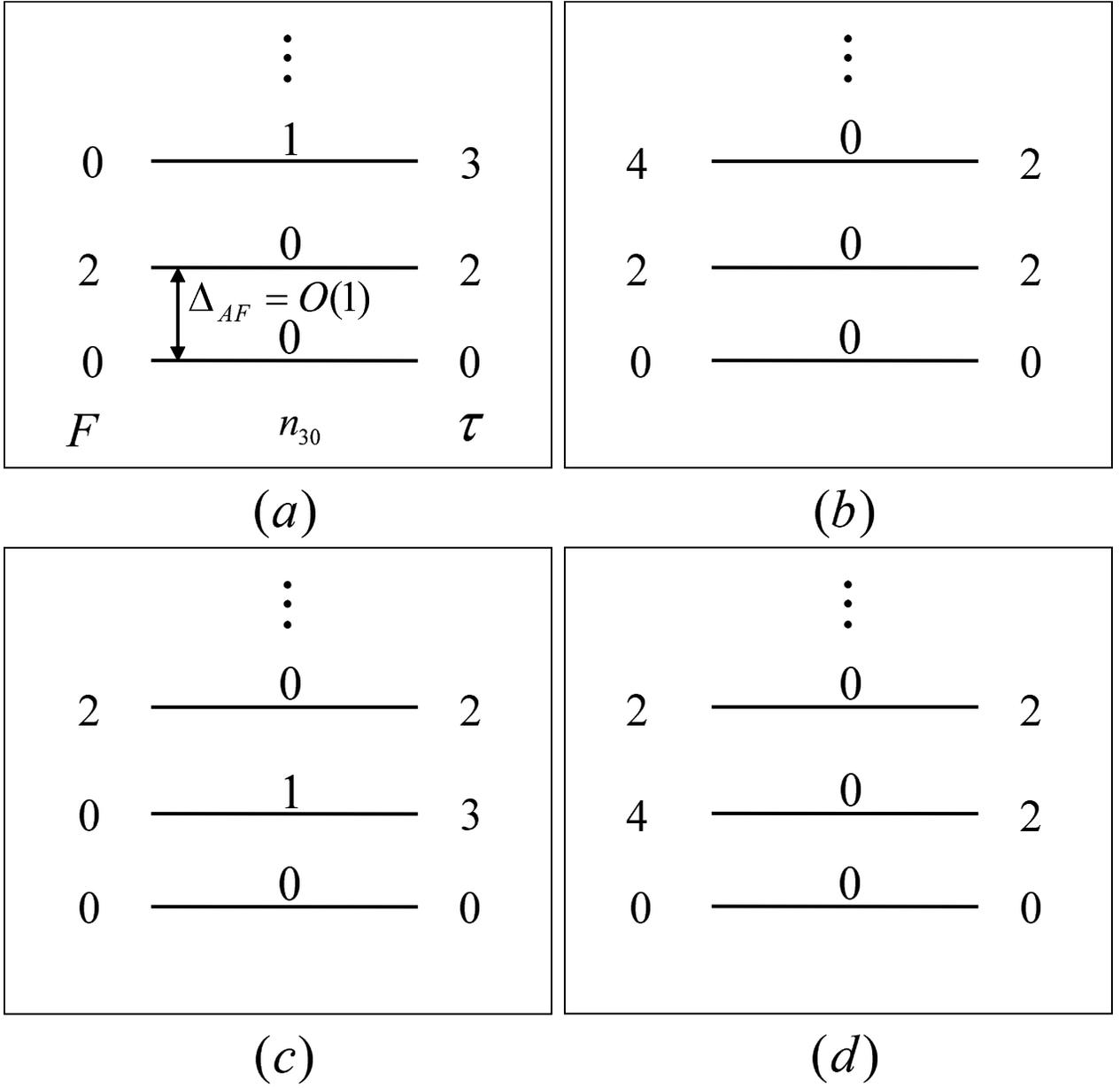}
  \caption{Low-lying eigenspectra of a spin-2 BEC for the antiferromagnetic phases with $N=2k$.
           The values of $F$, $n_{30}$, and $\tau$ of each state are shown to the left, middle, and right, respectively.
		   (a) $d_1>0$ and $\frac{3}{4}d_1<d_5<\frac{5}{2}d_1$. (b) $d_1>0$ and $d_5>\frac{5}{2}d_1$.
		   (c) $d_1>0$ and $0<d_5<\frac{3}{4}d_1$. (d) $d_1<0$ and $d_5>-\frac{4N+2}{N+3}d_1$.
           $\Delta_{AF}$ denotes the energy gap of the antiferromagnetic ground state.}
  \label{Fig.3}
 \end{center}
\end{figure}

\begin{figure}
 \begin{center}
  \includegraphics{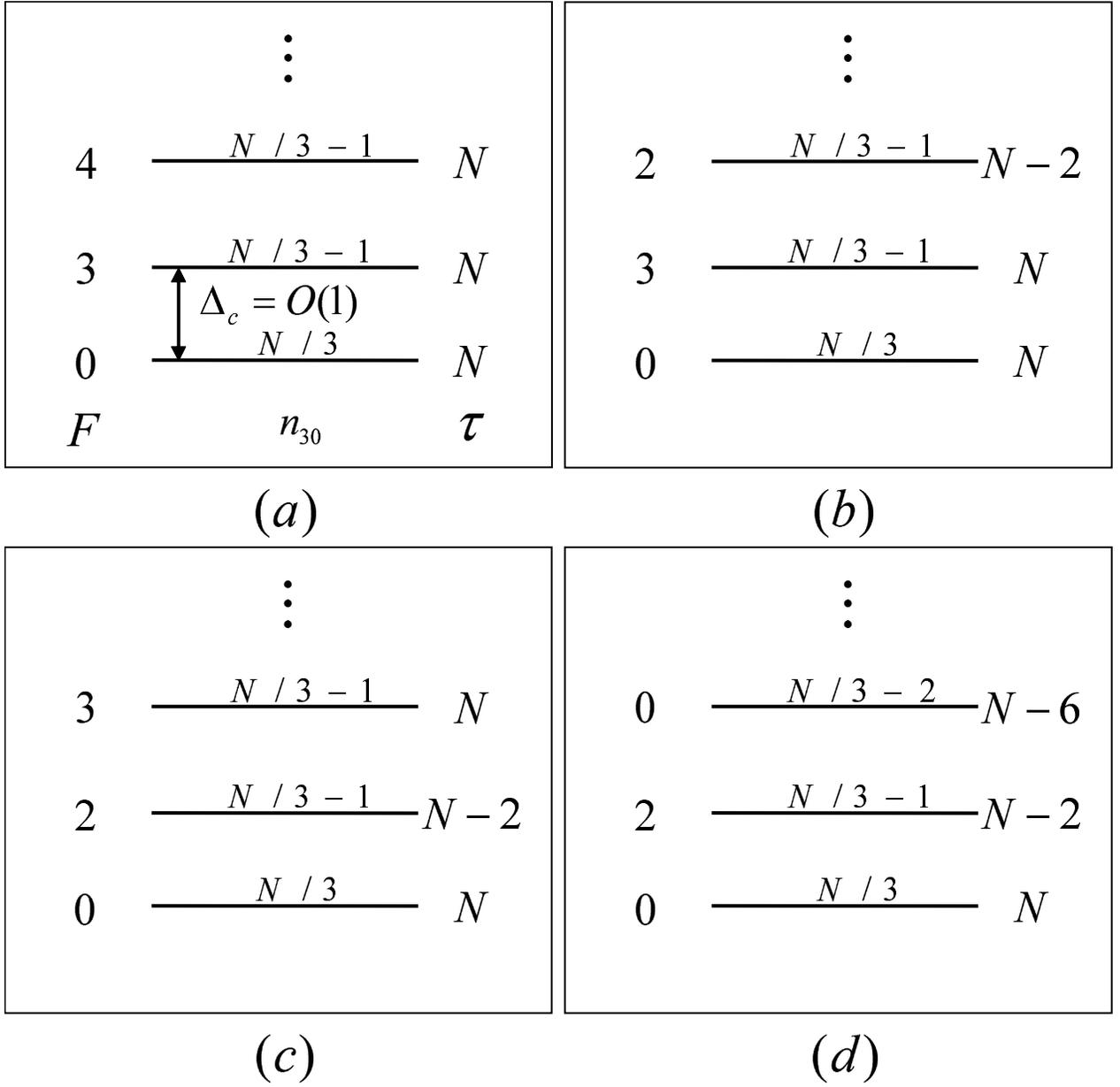}
  \caption{Low-lying eigenspectra of a spin-2 BEC for the cyclic phases with $N=3k$ and $d_1>0$. 
           The values of $F$, $n_{30}$, and $\tau$ of each state are shown to the left, middle, and right, respectively.
		   (a) $d_5<-\frac{7}{2N+1}d_1$. (b) $-\frac{7}{2N+1}d_1<d_5<-\frac{3}{2N+1}d_1$.
		   (c) $-\frac{3}{2N+1}d_1<d_5<-\frac{2}{2N-3}d_1$. (d) $-\frac{2}{2N-3}d_1<d_5<0$.
           $\Delta_{C}$ denotes the energy gap of the cyclic ground state.}
  \label{Fig.4}
 \end{center}
\end{figure}

This leads to a marked difference with the spin-1 case because the low-lying excitation spectra of a spin-1 BEC are uniquely determined when
the ground-state phases are specified. 
In Fig. \ref{Fig.2}, \ref{Fig.3}, and \ref{Fig.4}, we give diagrams of the low-lying eigenspectra up to
the second excited states in the ferromagnetic phase with $N=2k$, antiferromagnetic phase with $N=2k$, and cyclic phases with $N=3k$, respectively.
The other particle-number cases can also be found in a similar way and a rich variety of cases appear in considering the higher excitation.
In the ferromagnetic case, while the diagram (a) disappears in the limit $N\to\infty$, 
the other diagrams do not disappear in this limit. Similarly, while in the antiferromagnetic phase all diagrams do not disappear,
in the cyclic phase the only remaining diagram is (a) in this limit.   
Since $\displaystyle \Big|\frac{d_1}{d_5}\Big|$ is of order of $1\sim 10$ for spin-2 species \cite{ciobanu,klausen,kempen}, the low-lying energy-level spacing of the antiferromagnetic and cyclic phases are of order of 1, whereas 
that of ferromagnetic phase is of order of $N$. Considering the huge degeneracies of the ground states found in the antiferromagnetic and cyclic phases \cite{ueda},
it is of interest to study the response of these ground states to symmetry-breaking perturbations. The investigation of this problem is underway.

\section{Summary and discussions}
In this paper, we have shown that dynamical symmetry completely determines the exact eigenspectra and eigenstates of spin-1 and spin-2 BECs 
in the single-mode approximation. In particular, a spin-2 BEC in this approximation corresponds to the $U(5)$ limit of the interacting boson model of atomic nuclei. 
We have also shown that dynamical symmetry alone cannot solve the eigenvalue problem in the spin-3 case, because the Casimir operator
of the exceptional group $G_2$ is proportional to that of the $SO(7)$ group. We have, however, identified the class of the coupling constants 
for which the exact eigenspectrum can be found by dynamical symmetry alone. 
 
Compared with the spin-1 case, the new term that appears in the spin-2 case is the term $\displaystyle\hat{\mathcal{P}}_0$ which reflects the $SO(5)$ symmetry in the spin-2 case. 
If we put $d_1=p=0$ in Eq. (\ref{spin2}), we find that the symmetry of the Hamiltonian is not $U(1)\times SO(3)$ but $U(1)\times SO(5)$. 
We can generalize this fact to the spin-$f$ case because $\displaystyle\hat{\mathcal{P}}_0 $ is written as
\beq
\hat{\mathcal{P}}_0 =\frac{1}{2f+1}\left[ \hat{N}(\hat{N}+2f-1)-\hat{C}_{2}(SO(2f+1))\right] .
\eeq
That is to say, the Hamiltonian has the $U(1)\times SO(2f+1)$ symmetry within a specific class of coupling constants.

The eigenspectrum of a spin-1 BEC in Eq. (\ref{spin1}) is characterized only by the total number of particles and the total angular momentum, 
which reflect the gauge invariance and the isotropy of space respectively. The eigenspectrum of a spin-2 BEC in Eq. (\ref{spin2}) involves 
an additional quantum number $\tau$, which does not reflect either the gauge or space-time symmetry
but arises from the Casimir operator of the $SO(5)$ group;
$\tau$ denotes the number of particles not in pairs of $F=0$.
The missing label $n_{30}$, which describes the number of spin-singlet trios,
appears only in the eigenstate (\ref{state2}) and not in the eigenspectrum (\ref{eigen2}) of a spin-2 BEC.
Within the specific class of the parameters with $e_3=0$, 
the eigenspectrum of a spin-3 BEC is completely characterized by the same set of quantum numbers as that of a spin-2 BEC, that is, $N$, $F$, $F_z$, and $\tau$,
and it is in this class that the eigenvalue problem of a spin-3 BEC can be solved by the dynamical symmetry alone.

We have also discussed the low-lying eigenspectra and eigenstates of spin-1 and spin-2 BECs in the absence of external magnetic field.
For the spin-1 case, the phase boundary between the ground states with (\ref{state1}) and (\ref{eigen1}) corresponds to
that of mean-field theory.
The exact low-lying excitation spectra are uniquely determined by the sign of $c_1$.
On the other hand, for the spin-2 case, the phase boundaries between the ground states with (\ref{state2}) and (\ref{eigen2})
are different from those of the mean-field theory.
These differences disappear in the limit $N\to\infty$ except for the phase boundary within the antiferromagnetic phase with $N=2k+1$.
However, since all experiments of BECs in the cold atomic gases are done for a finite number of atoms,
these differences seem to be real issues.
In addition, the low-lying excitation spectra cannot be specified even if the ground states are specified,
and the coupling-constant dependence of the low-lying excitation spectra is stronger than those of the ground states.
This is because while the quantum number which means the number of particles not in pairs of $F=0$ is equal to the quantum number $F$ for a spin-1 BEC, 
it is not so for the spin-2 case in which there arises an additional quantum number $n_{30}$ which connect these quantum numbers. 
We also show that for the case of spin-1 and spin-2 BECs, the low-lying energy-level spacings of antiferromagnetic and cyclic phases
are by a factor of $N$ smaller than those of the ferromagnetic phase. 
This means that for the antiferromagnetic and cyclic phases quasi-degenerate spectra emerge which may cause symmetry-breaking transitions to yet unexplored many-body states.

Note added.-- At the time of submission of this paper, we became aware of a paper \cite{heinze} by Van Isacker and Heinze
who discuss the exact ground-state phase structure with arbitrary spin in the absence of external magnetic field.
\begin{acknowledgments}
S.U. thanks Prof. T. Hatsuda, Dr. N. Shimizu, Dr. S. Sasaki, and Mr. T. Kanazawa for helpful comments. 
\end{acknowledgments}

\appendix*
\section{recoupling formulae}
The Wigner 6-$j$ symbol is defined by \cite{iachello}
\beq
\begin{Bmatrix} j_1 & j_2 & j_3 \\m_1 & m_2 & m_3 \end{Bmatrix}
=\sum_{\mu_1,\mu_2\mu_3,\nu_1,\nu_2,\nu_3}(-1)^{j_1+j_2+j_3+m_1+m_2+m_3+\mu_1+\mu_2+\mu_3+\nu_1+\nu_2+\nu_3} \nonumber\\
\times\begin{pmatrix} j_1 & j_2 & j_3 \\ \mu_1 & \mu_2 & \mu_3 \end{pmatrix}\begin{pmatrix} j_1 & m_2 & m_3 \\ -\mu_1 & \nu_2 & -\nu_3 \end{pmatrix}
\begin{pmatrix} m_1 & j_2 & m_3 \\ -\nu_1 & -\mu_2 & \nu_3 \end{pmatrix}\begin{pmatrix} m_1 & m_2 & j_3 \\ \nu_1 & -\nu_2 & -\mu_3 \end{pmatrix},
\eeq
where $\displaystyle \begin{pmatrix} j_1 & j_2 & j_3 \\ \mu_1 & \mu_2 & \mu_3 \end{pmatrix}$ is the Wigner 3-$j$ symbol and is defined by
\beq
\begin{pmatrix} j_1 & j_2 & j_3 \\ \mu_1 & \mu_2 & \mu_3 \end{pmatrix}=\frac{(-1)^{j_1-j_2-\mu_3}}{\sqrt{2j_3+1}}<j_1\mu_1j_2\mu_2|j_3,-\mu_3>.
\eeq 
The Wigner 6-$j$ symbol is related to the Wigner 3-$j$ symbol by \cite{iachello}
\beq
\sum_{m_3}\begin{pmatrix} j_1 & j_2 & j_3 \\m_1 & m_2 & m_3 \end{pmatrix}\begin{pmatrix} j_1^{'} & j_2^{'} & j_3 \\m_1^{'} & m_2^{'} & -m_3 \end{pmatrix}
=\sum_{j_3^{'},m_3^{'}}(-1)^{j_3+j_3^{'}+m_1+m_1^{'}}(2j_3^{'}+1)\nonumber\\
\times \begin{Bmatrix} j_1 & j_2 & j_3 \\j_1^{'} & j_2^{'} & j_3^{'} \end{Bmatrix} 
\begin{pmatrix} j_1^{'} & j_2 & j_3^{'} \\m_1^{'} & m_2 & m_3^{'} \end{pmatrix}\begin{pmatrix} j_1 & j_2^{'} & j_3^{'} \\m_1 & m_2^{'} & -m_3^{'} \end{pmatrix}. \label{recouple}
\eeq
Using Eq. (\ref{recouple}), one can derive the following recoupling formulae: \\
For the spin-1 case,
\beq 
[\hat{a}^{\dagger}\times \hat{\tilde{a}}]^{l}\cdot[\hat{a}^{\dagger}\times\hat{\tilde{a}}]^{l}=(2l+1)\sum_{l^{'}}
\begin{Bmatrix} 1 & 1 & l^{'} \\1 & 1 & l \end{Bmatrix}
[\hat{a}^{\dagger}\times\hat{a}^{\dagger}]^{l^{'}}\cdot[\hat{\tilde{a}}\times\hat{\tilde{a}}]^{l^{'}} 
+\frac{2l+1}{3}\hat{N};\label{rec1}
\eeq
For the spin-2 case,
\beq 
[\hat{a}^{\dagger}\times \hat{\tilde{a}}]^{l}\cdot[\hat{a}^{\dagger}\times\hat{\tilde{a}}]^{l}=(2l+1)\sum_{l^{'}}
\begin{Bmatrix} 2 & 2 & l^{'} \\2 & 2 & l \end{Bmatrix} 
[\hat{a}^{\dagger}\times\hat{a}^{\dagger}]^{l^{'}}\cdot[\hat{\tilde{a}}\times\hat{\tilde{a}}]^{l^{'}} 
+\frac{2l+1}{5}\hat{N};\label{rec2}
\eeq
For the spin-3 case,
\beq 
[\hat{a}^{\dagger}\times \hat{\tilde{a}}]^{l}\cdot[\hat{a}^{\dagger}\times\hat{\tilde{a}}]^{l}=(2l+1)\sum_{l^{'}}
\begin{Bmatrix} 3 & 3 & l^{'} \\3 & 3 & l \end{Bmatrix} 
[\hat{a}^{\dagger}\times\hat{a}^{\dagger}]^{l^{'}}\cdot[\hat{\tilde{a}}\times\hat{\tilde{a}}]^{l^{'}} 
+\frac{2l+1}{7}\hat{N}.\label{rec3}
\eeq

\end{document}